\documentstyle[11pt]{article} 
\def\bas\def\baselinestretch{1.2} 
\input{psfig} 
\begin{document} 
\begin{flushright}  
{OUTP-00-29-P}\\   
\end{flushright}  
\vskip 2 cm 
\begin{center} 
{\Large {\bf A note on the RG flow in ($N=1$, $D=4$) Supergravity   \\
\vspace{0.35cm}
and applications  to $Z_3$ orbifold/orientifold compactification
\bigskip }} 
\\[0pt] 
\bigskip {\large 
{\bf Dumitru M. Ghilencea\footnote{
{{ {\ {\ {\ E-mail: D.Ghilencea1@physics.oxford.ac.uk}}}}}} } 
and {\bf Graham G. Ross\footnote{
{{ {\ {\ {\ E-mail: G.Ross1@physics.oxford.ac.uk}}}}}}
\bigskip }}\\[0pt] 
{\it Department of Physics, Theoretical Physics, University of 
Oxford}\\[0pt] 
{\it 1 Keble Road, Oxford OX1 3NP, United Kingdom}\\[0pt] 
\bigskip 
\vspace{3cm} Abstract
\end{center} 
{\small We apply the standard approach of RG flow for the gauge
couplings in N=1 D=4 Supergravity  to show how to  match its results
with  the  heterotic $Z_3$ orbifold  and 
Type IIB ${Z_3}$  orientifold-based models. 
Using only supergravity, anomaly cancellation
and the requirement of unification we determine the part of 
the  K\"ahler potential of the model  invariant under the
symmetries of the model. For  heterotic
orbifolds/type IIB orientifolds,  this shows that the lowest order
K\"ahler term of the dilaton has the  
structure $-\ln(S+{\overline S})$ in agreement with string 
calculations.
The structure of the holomorphic coupling is also 
found from arguments of unification and 
anomaly cancellation under the conjectured $SL(2,Z)_{T_i}$ 
symmetries of the models. A consequence of the latter is that 
in the case of the $Z_3$ orientifold the holomorphic coupling
necessarily contains a part with coefficient   proportional
to the one loop beta function, in agreement with string
calculations which do not however assume this symmetry. 
This gives circumstantial evidence for the existence of this symmetry 
at string level in $Z_3$ orientifold.
Finally, we comment on the values of the unification
scale and examine  the possibility of mirage unification
in which the effective unification scale may be situated far above
the string scale. 
} 
\newpage  
\setcounter{page}{1}
\setcounter{footnote}{0}

\section{Introduction} 
Considerable interest has recently been attracted by the initial
investigation  \cite{ibanez1} of sigma-model symmetries in 
Type II B D=4, N=1 orientifolds at the level of the effective
Lagrangian.
The motivation for the study of such symmetries is suggested by
the case of D=4, N=1 heterotic orbifold models. There an
$SL(2,R)_{T_i}$  transformation leaving invariant the classical 
Lagrangian also corresponds to an exact $SL(2,Z)_T$ symmetry of the
underlying string ($T$ duality). What do we know on the
orientifold side?  The proposed 
``strong-weak'' duality \cite{polchinski}
between the heterotic  and type I  vacua
in ten dimensions suggested the existence
of links between heterotic and type I models in lower dimensions
\cite{angelantonj1}. This has lead in particular to a generalised
heterotic-type II B orientifold duality in four dimensions in which
both models are weakly coupled \cite{angelantonj1} for some regions 
of the moduli space.
%% expectation that a mechanism for 
%% implementing such symmetries and the cancellation of their anomalies 
%% induced at the quantum level might exist in the case
%% of type II B orientifolds as well. 
%% This  duality, which is of the weak coupling - strong coupling 
%% type in ten dimensions, upon  compactification to lower dimensional 
%% space-time gives rise to 
%% weak coupling -  weak coupling duality for some regions of the moduli space 
%% \cite{angelantonj1}.  
%% This further  suggests that it may  be possible to have
%% dual models which are simultaneously weakly coupled.
An example of this latter duality  is provided by $Z_3$
orbifold-orientifold models \cite{angelantonj1,nilles0} without Wilson
lines and these are the  models we will consider in the following.
Their very similar spectrum strengthens the suggestion that they
are indeed dual.
The presence of the sigma model symmetry and its anomaly cancellation
in  the $Z_3$ heterotic case thus suggests its existence and
associated anomaly cancellation in  the $Z_3$ orientifold model as
well. Thus studying the anomaly cancellation on the orientifold side 
may help us find out more about the proposed ``weak-weak'' 
duality in four dimensions.
Here we examine the consequences of this $SL(2,Z)_T$ symmetry in the $Z_3$
orientifold model and  discuss its implications for the RG flow of the
gauge couplings, intimately connected to its anomaly cancellation.
Previous analyses and associated difficulties 
of $SL(2,Z)_{T}$ symmetry on  the orientifold side were discussed in
\cite{ibanez1,nilles0,klein,nilles,lavignac,ibanez4,antoniadis}.

In the case of the heterotic string, the models based on the 
$Z_3$ orbifold have the gauge group  $SU(12)\times SO(8)\times
U(1)_A$ where $U(1)_A$ is anomalous. This anomaly, which is universal,
 is cancelled  by the shift of the dilaton. The Fayet
Iliopoulos term, dilaton dependent, is cancelled by shifts of some
additional fields. These are the so-called ``blowing up''
modes of the orbifold which are charged only under the 
$U(1)_A$ and compensate the Fayet Iliopoulos term  when they acquire some
expectation value, without breaking further the gauge group.
The mass of the anomalous gauge boson becomes equal to the string scale
and decouples from the RG flow of the model.

For the case of the orientifold  (we mainly refer to the  
$Z_3$ orientifold, but
some aspects in the following are more general) 
the situation is slightly different.
In this case the gauge group is also $SU(12)\times SO(8)\times
U(1)_A$ where $U(1)_A$ is again anomalous. However, in this 
case the anomalies
are gauge group dependent and may be cancelled by a generalised Green
Schwarz mechanism. This can be realised by the shifting
\cite{ibanez1,ibanez2,nilles0} of the twisted axions 
which are singlets under 
the gauge group. The anomalous vector superfield mixes with these 
states and acquires a mass which is again of string scale order, 
regardless of the particular 
choice of the K\"ahler potential for the twisted moduli \cite{nilles0}.

After determining that the two models  have
a similar spectrum and the same gauge group below the string scale, 
it is interesting to analyse whether the weak coupling - weak
coupling  duality really holds beyond tree level. This was the 
purpose of the 
investigation in  \cite{nilles} at the one loop level.  Starting from the 
RG flow in the string (linear) basis and using the
linear-chiral duality relation \cite{antoniadis} it was shown 
\cite{nilles} that the  results for the low energy couplings 
raise doubts about the  existence of the $Z_3$ orbifold-orientifold 
duality \cite{nilles}. There is however a difficulty in analysing 
the relation between the couplings of the 
 $Z_3$ orbifold and $Z_3$ orientifold models.
This is due to the fact that the string scale in the latter case was not
invariant under the proposed $SL(2,Z)_{T_i}$ symmetry. Moreover, 
in the orientifold limit, this means that
 the low energy physics, as it emerges from
the linear basis formula (\ref{eq1}) \cite{antoniadis} is
not  invariant under the symmetry. In the linear basis
\cite{antoniadis}
for the $Z_N$ orientifold
\begin{equation}\label{eq1}
g_a^{-2}(\mu)= l^{-1}+\sum_{k=1}^{(N-1)/2}
 s_{ak} m_k+ \frac{b_a}{8\pi^2} \ln\frac{M_I}{\mu}
\end{equation}
where $l$ is the string coupling, $s_{ak}$ is a coefficient 
proportional to one loop beta function $b_a$ and $m_k$ is
the scalar component of the twisted linear multiplet describing the
blowing up modes of the associated orbifold.
In the following we will drop the index $k$ whenever we refer to 
the particular case of the $Z_3$ orientifold.
The orientifold limit $m_k\rightarrow 0$ together with the 
dependence of the tree level string scale 
$M_I$ on the $T_i$ moduli\footnote{At string tree level 
this is given by $M_I^2=(Re S/\prod_{i=1}^{3} 
Re T_i)^{1/2} M_P^2/ (2 Re S)$.}
leads to the conclusion that $g_a(\mu\sim M_z)$ changes  
under $SL(2,Z)_{T_i}$. 
This is obviously inconsistent with the fact that 
 $g_a(\mu\sim M_z)$ is a physical 
observable and it must remain invariant under this transformation
if it is a symmetry of the theory. 

There are three possible
resolutions to this problem. 
The first is that  there may be additional terms 
in eq.(\ref{eq1}) which  render $g_a(\mu\sim M_z)$ invariant. 
% although this seems to be unlikely \cite{nilles} on anomaly 
% cancellation considerations.
The second possibility is that the
string scale definition to be used in (\ref{eq1}) is changed from 
the tree level value $M_I$ and its
one-loop improved value is  invariant under the 
transformation of $T_i$. This is similar to the heterotic case where
the tree level string scale $M_H$ depends on $S$, $M_H^2\sim
M_P^2/(S+\overline S)$; the dilaton 
changes under $SL(2,Z)_{T_i}$ (for universal anomaly cancellation) so
$M_H$ is not invariant, but
the one-loop improved heterotic string scale is invariant 
under this symmetry, as we will discuss later.
The last possibility is that the
proposed symmetry $SL(2,Z)_{T_i}$ is not there at all in the 
case of $Z_3$ orientifold models, putting into doubt the 
existence of the proposed $Z_3$ orbifold-orientifold duality.

In this work we investigate these possibilities adopting 
a different perspective.  We work only in
the chiral basis of $N=1$, $D=4$ Supergravity and do not use the
(questionable) linear-chiral multiplet duality relation \cite{antoniadis}.
%  \footnote{To match the string result (\ref{eq1}) one should actually
%  consider the RG flow in the  linear basis in Supergravity and then 
%  perform a transformation to the  chiral basis.}.  
We assume there is an effective supergravity theory
below some cut-off scale $\Lambda$. As we will show, 
this approach provides relevant information for anomaly cancellation 
and the unification of the gauge couplings. 
This is interesting because it will give us
some understanding as to whether the proposed mirage unification
scenario  initially suggested by \cite{ibanez4}
can be made to work for $Z_3$ orientifolds. 
This requires a careful investigation of the RG flow for
both models based on the $Z_3$ orbifold/orientifold.

 The RG flow in 
$N=1$ $D=4$ Supergravity models was introduced in \cite{kl1} (see
also  \cite{kl2,kl3} for applications to the heterotic case) and we
will briefly review some aspects of it in the next section, with
emphasis on the importance of the various terms contributing to the
(perturbatively exact) 
running. A more illustrative form of the RG flow is  presented 
in Section \ref{sectionrge} which makes manifest how string theories 
% (with a low energy stage of supergravity) 
renormalise, through the K\"ahler
terms for the moduli fields, the bare coupling and
 the (high) scale to which the effective (canonical) gauge couplings run.
This discussion will show that the assumption there is gauge coupling
unification 
in supergravity models suggests the existence of a link at a deeper
level in string theory between the structure of the K\"ahler potential
for moduli fields (other than the dilaton) and that of the dilaton
itself.  In Section \ref{orbifold}  we discuss the case of the $Z_3$
orbifold.
We stress the importance of the one loop improved string scale 
which emerges
as the natural cut-off in the RG flow of the (effective) 
gauge couplings and which
should be used in models which want to test the  
weak coupling - weak coupling duality, at one-loop level.
In Section \ref{orientifold}  we perform a similar analysis to 
investigate the
gauge couplings running in the case of $Z_3$ orientifold model. 
Using only anomaly cancellation
arguments in $D=4$, we  recover on purely field theoretic grounds,
 the result from string theory \cite{antoniadis,ibanez1} 
that the coefficient of the twisted moduli entering the RG flow is 
indeed proportional to the one loop beta
function for the case of $Z_3$ orientifold. 
The calculation also shows that the effective unification scale 
could be   far above the string scale (mirror unification), 
perhaps  close to the Planck scale. 
Unfortunately our analysis cannot say more about this
value since the  v.e.v. of  the twisted moduli is not determined
by the Fayet Iliopoulos mechanism. This  is due to our  
present lack of understanding of the linear-chiral duality
relation, which relates the string result (\ref{eq1})  to that of  
Supergravity RG flow for  $g_a(\mu\sim M_z)$, the question of invariance 
of the string scale under the assumed  $SL(2,Z)_{T_i}$ symmetry
and the cancellation of anomalies in type IIB orientifolds \cite{nilles}. 
These  issues  may affect  the results obtained from the Fayet
Iliopoulos mechanism in the orientifold limit. 
This mechanism has been investigated in ref.\cite{klein} 
but its conclusions seem to be in conflict with the proposed
linear-chiral multiplet duality relation \cite{antoniadis}
suggesting further study is necessary.
  
To avoid some of these problems for the $Z_3$ orientifold models 
we present the 
results in terms of an unknown function ${\cal G}$ invariant under the
$SL(2,Z)_{T_i}$ symmetry and speculate about its effects on
the value of the  unification scale.  Our conclusions are
presented in the last section.

\section{RG flow, symmetries and the link with string theory}
The standard procedure for analysing the RG flow equations in local
supersymmetric theories and their matching with string
theory\footnote{for the heterotic case} 
was introduced by Kaplunovsky and Louis  
in ref.\cite{kl1,kl2} (see  also \cite{kl3}). 
Its application to string models in general requires a
knowledge of the  K\"ahler potentials of the moduli fields as well 
as the structure of the  holomorphic couplings. 
 
Let us consider the link between the supergravity RG flow 
and string theory - in our case heterotic orbifolds
and Type II B orientifolds. There are two important  differences 
between these cases.
Firstly, the ultraviolet  cut-off $\Lambda$ of the RG flow 
of the Wilsonian coupling in the effective theory of supergravity may not 
be  the same in both cases. However, whatever its value may be, 
by {\it definition} \cite{kl1}
it must be independent of all fields/moduli \cite{kl1,kl2}. 
From the point of view of the effective supergravity theory
$\Lambda$ must be constant in Planck units, and 
it can therefore be set equal to the Planck scale $M_P$.
This applies to the effective supergravity models  of the 
heterotic strings \cite{kl1,kl2}. It is not clear to us 
whether this should also be the case  for the type I string models we 
consider here, although certainly $\Lambda$ 
must be a moduli/field independent quantity.
Secondly, for the two classes of string models we consider,
the expressions for the holomorphic function $f_a$ and the 
dependence on the  moduli fields of the  K\"ahler 
potentials may be different. 
Their input from string theory in both cases
realises the connection between effective supergravity and string theory, 
allowing a test of string unification at (N=1, D=4) supergravity level.
However, throughout this paper, we will  keep the string input 
at minimum and 
use only supergravity arguments for most of the calculations.
A great deal of the structure of the Wilsonian coupling and even their
relationship to the K\"ahler moduli can be understood on pure
supergravity grounds together with gauge unification  which we
assume applies. This will be detailed in the following sections.
 
The integral of the RG flow for the effective gauge 
couplings $g_a$ in rigid supersymmetry 
follows from the exact beta function 
for $g_a$ introduced by Novikov, Shifman, Vainshtein, Zakharov 
(NSVZ), \cite{nsvz,hamed}. The result after integration is 
\begin{equation}\label{lab1}
F_a=g_a^{-2}(\mu)+\frac{b_a}{8\pi^2} \ln \mu-\frac{T(G_a)}{8\pi^2}\ln
g_a^{-2}(\mu) + \sum_{r} \frac{T_a^{(r)}}{8 \pi^2} \ln  Z_r
(\rho,\mu) 
\end{equation}
where condition $d F_a/ d\mu=0$ recovers the ``NSVZ'' beta function.
Here $b_a=-3 T_a(G)+\sum_{r} T_a^{(r)}$.
After applying this equation at two different scales, one recovers 
the running  of the couplings between these scales to all 
orders in perturbation theory. The coefficients $Z$ are normalised to
unity at some arbitrary scale $\rho$. At two loop level one can 
reproduce the structure of the 
familiar RG flow of the Minimal Supersymmetric 
Standard Model (MSSM).

For the Wilsonian couplings of the {\it local} supersymmetric theories
we have the following  integral of the renormalisation group
\begin{equation}\label{lab2}
F_a=Re f_a + \frac{b_a}{8 \pi^2}\ln\Lambda + \frac{c_a}{16 \pi^2}{\cal
K}, \,\,\,\,\,\,\,\,\,\,\, {\cal K}=\kappa^2 K
\end{equation}
where $\Lambda$ is the ultraviolet cut-off scale. 
%  which by {\it definition} does not depend on either the moduli or  the 
%  so-called Weyl compensator fields \cite{kl1}.
Equation (\ref{lab2})  simply reflects that Wilsonian 
couplings do not renormalise beyond one-loop.
The last term in (\ref{lab2}) accounts for the super-Weyl anomaly
induced by the rescaling of the metric necessary to separate 
gravity from quantum field theory effects in the 
supergravity action \cite{kl1,bagger}. 
% Technically this anomaly is  caused by the
% fact that there is no field independent cut-off that respects the 
% super-Weyl symmetry and this is taken care of by the Weyl 
% compensator fields.
We use the notation
$c_a=-T_a(G)+\sum_{r}T_a^{(r)}$, 
and  $\kappa^2=8 \pi/M_P^2$. $K$ is the 
full K\"ahler potential which includes a 
term for moduli fields and one for the charged matter fields
\cite{kl1}. Its expression is given by
\begin{equation}\label{moduli}
K=K_{mod}+K_{matter}=k^{-2} \tilde {\cal K}+K_{matter}
\end{equation}
where the moduli part of the K\"ahler potential
$K_{mod}\sim {\cal O}(M_P^2)$ is the only relevant part in 
(\ref{lab2}), the charged matter term $K_{matter}\sim Z {\overline Q} Q$
being strongly suppressed relative
to the moduli contribution. Therefore 
${\cal K}\approx {\overline {\cal K}}$. Equations (\ref{lab1}),
(\ref{lab2}) above set the integral of the RG flow in models  with local
supersymmetry in the following form \cite{kl1,kl2,kl3} 
\begin{equation}\label{lab3}
g_a^{-2}(\mu)=Re f_a+\frac{b_a}{16 \pi^2}\ln \frac{\Lambda^2}{\mu^2}
+ \frac{1}{16 \pi^2}\left\{c_a {\cal K} +2 T(G_a) \ln g_a^{-2}(\mu)-2
\sum_{r} T_a^{(r)} \ln Z_{r}(\rho,\mu)\right\}
\end{equation}
A comment on the structure of equation (\ref{lab3}) and the way it is
linked to string theory is in place here. Although this discussion
is made in the context of the weakly coupled heterotic models, 
it is in fact  more general. Usually the link with string 
theory is established 
by taking the K\"ahler potential for the dilaton and other
moduli  as  given to some order of perturbation in 
string theory. Consider, for example, the
K\"ahler potential for the dilaton 
$ {\cal K} \sim -\ln (S+\overline S)$. Its contribution in the 
curly braces of (\ref{lab3})  is usually split into two terms, 
one given by
$-b_a/(16 \pi^2) \ln(S+\overline S)$ 
which combines with the cut-off $\Lambda$ to give
% \footnote{For 
% weakly coupled heterotic strings $\Lambda=M_P$ \cite{kl2}.}
the heterotic string scale, $M_H^2\sim \Lambda^2/(S+\overline S)$,
while the remaining part $-(c_a-b_a)/(16 \pi^2)\ln(S+\overline S)$ 
combines with  $\ln g_a^{-2}(\mu)$ to give $2 T_a(G)\ln (S+\overline
S)^{-1}/g^{2}(\mu)$ in the curly braces. Further, this last
 term (pure gauge) is usually ignored
as being considered a higher order (two loop and beyond) term
in the effective field theory.
This is not correct because it is inconsistent to split the logarithmic
dependence of the dilaton into two pieces and to retain only $b_a \ln
(S+\overline S)$ while ignoring the part proportional to 
$2 T_a(G)\ln(S+\overline S)$.
Since the string expansion parameter is $\sim 1/Re S$ 
truncating such an expansion  to keep 
only part of $\ln(S+\overline S)$  term 
makes the whole  calculation only valid in 
${\cal O}(Re S)$ order i.e. string tree level.

The presence of the terms $Z_\alpha(\rho,\mu)$
is due to wavefunction renormalisation {\it at/above}
and {\it below} some cut-off\footnote{which turns out to be the string
scale, see later.} of the charged matter 
fields.  Usually only their value at/above the cut-off is retained
to give contribution to the so called one-loop string threshold
effects (dependent on the $T$ moduli). This contribution is due to
the K\"ahler terms for charged matter fields which are not
canonically normalised and comes with what can be considered as a 
wavefunction coefficient of string origin, 
$Z_r(\rho,\tilde \Lambda) \sim \prod_{i=1}^{3}
(T_i+\overline T_i)^{n_r}$ where 
$n_r$ are modular weights of the matter fields, $\rho$ is some
arbitrary scale\footnote{The scale $\rho$ is not physical, it simply
corresponds to a particular point in the moduli space where $Z$ is
normalised to unity.} and $\tilde \Lambda$ is the string scale. 
The contribution (computable in field theory) to $Z$ below the cut-off
(later we will  identify it with the string scale, $\tilde \Lambda$)
is needed to account for  matter and mixed gauge-matter contributions
to the  running of the ({\it effective}) couplings. 
 At one loop order there is 
a gauge contribution of the form $\ln Z(\tilde\Lambda,\mu)\sim \ln
g_a(\tilde\Lambda)/g_a(\mu)$. This together with the fact 
$g_a^2(\tilde\Lambda) \sim (S+\overline S)^{-1}$ shows it is of comparable 
magnitude to the dilaton contribution $\ln(S+\overline S)$
and must be included. A more elegant motivation for including
$\ln Z$ terms below the cut-off scale is due to the fact that they
originate from  anomaly  cancellation (from the rescaling of 
chiral superfields) just as the  terms $\ln g_a^{-2}$ do \cite{hamed}  
(from the rescaling of vector superfields), 
so they  are on equal footing. 

The effective couplings $g_a$ are physical quantities, 
and therefore they must be invariant under the 
symmetries of the theory.
In particular we require the invariance of the low 
energy physics  
under the K\"ahler symmetry  transformations of the K\"ahler potential
$K$ and of the superpotential $W$, given by 
($\phi$ stands for the moduli fields and $Q$ for the charged 
matter fields)  \cite{kl1}
\begin{eqnarray}\label{lab5}
&K(\phi',{\overline{\phi'}},Q',{\overline {Q'}})&\rightarrow
K(\phi,{\overline \phi},Q,{\overline Q})+J(\phi,Q)+
{\overline J}(\overline\phi,\overline Q)\\
&W(\phi',Q')&\rightarrow W(\phi,Q) exp(-\kappa^2 J(\phi,Q))
\end{eqnarray}
At the classical level, these transformations leave the action invariant 
\cite{{kl1},{kl2},{kl3}}. At the quantum level these symmetries
imply a transformation of the 
holomorphic coupling $f_a$ (K\"ahler anomaly) which is
shifted by a quantity proportional to $J(\phi)$ as  follows 
from (\ref{lab3}),(\ref{lab5}).
In a similar way, a symmetry transformation of the matter fields 
of the form 
\begin{equation}\label{lab4}
Q^r\rightarrow Q^{\prime r} = Y^{r}_{r'}(\phi)Q^{r'}
\end{equation}
induces another anomaly with  
$ Q^\dagger Z Q = Q^{' \dagger} Z' Q' $, so
$Z\rightarrow Z'=Y^{\dagger -1} Z Y^{-1}$. 
This transformation together with that of the wavefunction
coefficients  leaves the K\"ahler potential for charged matter 
fields invariant,
but at the quantum level this rescaling of the fields induces an
anomalous term.
For  low-energy physics (i.e. $g_a(\mu)$)  to stay invariant 
under the combined effect of transformations (\ref{lab5}) and (\ref{lab4})
one finds the following transformation for the holomorphic coupling
\cite{{kl1},{kl2},{kl3}} 
\begin{equation}\label{lab6}
f_a(\phi)\rightarrow f_a(\phi')=f_a(\phi)-\frac{c_a}{8\pi^2}
\kappa^2 J(\phi)-\sum_{r}\frac{T_a^{(r)}}{4 \pi^2}\ln Y^{(r)}(\phi)
\end{equation}
This is a very important relation as it 
relates  two values of the Wilsonian coupling $f_a$
at different points in the moduli space \cite{kl1}. 

\section{$N=1$ $D=4$ Supergravity and RG flow}\label{sectionrge}
From eq.(\ref{lab1}) evaluated at two different scales $\mu$
and $\mu'$ ($\mu \leq \mu'$) one finds to all orders in perturbation
theory that the effective couplings run as follows
\begin{equation}\label{mssmlike}
g_a^{-2}(\mu)=g_a^{-2}(\mu')+\frac{-3
T_a(G)}{8\pi^2}\ln\frac{\mu'}{\mu
\left({g_a^2(\mu')}/{g_a^2(\mu)}\right)^{1/3}}
+\sum_{r}\frac{T_a^{(r)}}{8\pi^2}\ln \frac{\mu'}{\mu
Z_r(\mu',\mu)}
\end{equation}
Here we have used the fact that $Z_r(\rho,\mu)=Z_r(\rho,\mu') 
Z_r(\mu',\mu)$. 
% with $Z_r(M,M)=1$.
In the above RG flow the second term on the r.h.s. 
is the pure gauge term (exact to all orders)
while the last term is the mixed matter-gauge and matter only (Yukawa)
term again exact to all orders in perturbation theory. 
Computing $Z$ at one loop,  eq.(\ref{mssmlike})  reproduces the 
structure of the well-known two-loop running in the MSSM in the presence of 
Yukawa couplings \cite{nsvz,unif1}. 

If the gauge couplings unify at some scale (in this case $\mu'$ which
also provides the cut-off) $g_a(\mu')=g_0$. 
Eq.(\ref{mssmlike}) gives the RG 
flow in rigid supersymmetry, while we know that the effective 
theory close to the string scale should actually be  that of 
{\it supergravity}. However, the phenomenological success of
eq.(\ref{mssmlike}) in the context of the MSSM in relating low energy
values of the gauge couplings suggests any corrections to it should be small.
It would therefore be good to have a 
deeper understanding, at the supergravity level,  of why 
eq.(\ref{mssmlike}) is such a good approximation and of how 
the cut-off $\mu'$ and the  bare coupling $g_0$ emerge.

In string theory the ratio of the gauge couplings is predicted. 
The effective  (N=1, D=4) supergravity theory leads to a  
test of such  string predictions using the  RG flow.
In this the gauge couplings unification condition must be imposed
as a boundary condition.

The RG flow in supergravity is similar to that in 
eq.(\ref{mssmlike}) as long as the space-time is nearly flat,
but at scales of order ${\cal O}(\mu')$   gravity effects become more 
important. At such scales
the running of the couplings is  given by (\ref{lab3}) which we 
may present in a a form similar to (\ref{mssmlike}) 
% to make obvious  the contribution to the RG flow
% of various sectors (gauge and matter) of the theory 
% as well as the bare coupling and the cut-off of the effective
% couplings. We have
\begin{equation}\label{sugra}
g_a^{-2}(\mu)= Re f_a+\frac{-3 T_a(G)}{8\pi^2}\ln
\frac{\Lambda e^{{\cal K}/2}}{\mu \left({e^{\cal K}}/
{g_a^2(\mu)}\right)^{1/3}} +\sum_{r}\frac{T_a^{(r)}}{8\pi^2}\ln 
\frac{\Lambda e^{{\cal K}/2}} {\mu Z_r(\rho,\mu)}
\end{equation}
This equation is very similar to (\ref{mssmlike}) in its structure, 
and leads to the interpretation of the  ``cut-off'' as  
$\Lambda e^{{\cal K}/2}$,  the scale present in the second and
third terms in the r.h.s. of 
equation\footnote{If we considered instead 
$\Lambda e^{w {\cal K}/2}$ as a cut-off and with $e^{v {\cal K}}$ 
in the denominator of the  second term in  the r.h.s.  of (\ref{sugra})
acting as a bare coupling ($w,v$ are arbitrary constants), 
the condition $b_a w+ 2 T_a(G) v=c_a$ gives $b_a(w-1)= 2 T_a(G)(1-v)$
with (unique) gauge group independent solution $w=v=1$. This motivates our 
interpretation of $\Lambda e^{{\cal K}/2}$ as a potential 
``cut-off'' in eq.(\ref{sugra}).}  (\ref{sugra}).
Comparing (\ref{sugra}) and (\ref{mssmlike})
we  see that  the second term on the r.h.s. of
(\ref{sugra}) is the (all orders) pure gauge  contribution 
while the last term is the matter contribution (again to all orders in 
perturbation theory). The presence of the ``cut-off''
$\Lambda e^{{\cal K}/2}$ instead of $\Lambda$ 
%(as in eq.(\ref{lab2})) 
may be interpreted as a gravitational
effect corresponding  to the rescaling of the metric \cite{kl1}. 
The only effect of such a rescaling is to change the scale up 
to which the effective couplings 
run as compared to the holomorphic coupling  running, eq.(\ref{lab2});
the structure of the running  is similar in both cases,
eqs.(\ref{mssmlike}), (\ref{sugra}).
Moreover, evaluating (\ref{sugra}) at two different scales
$\mu$ and $\mu'$ and subtracting, we recover
eq.(\ref{mssmlike}) which corresponds to flat space-time RG flow, 
strengthening  our assertion that only the cut-off  of the running
and the bare coupling (i.e. the boundary conditions) 
change when going to the effective supergravity case. 
Further, since the leading contribution in ${\cal K}$ is that
of moduli fields\footnote{For this see eq.(\ref{moduli}).} which 
are intrinsically of string origin,
eq.(\ref{sugra}) makes manifest how string theory normalises the 
cut-off and the bare coupling of the effective supergravity theory through
the K\"ahler potential of moduli fields. 

The  unification condition for the gauge couplings 
is of the following form $Re f_a = e^{-{\cal K}}
= g_a^{-2}(\Lambda e^{ {\cal K}/2})$ up to scheme dependence terms. 
If this condition is  met, then eq.(\ref{sugra}) is of the same form
as eq.(\ref{mssmlike}). Thus eq.(\ref{sugra}) together with the gauge
unification requirement provides an explanation why corrections close
to the string scale preserve the successful (rigid supersymmetry) form
of eq.(\ref{mssmlike}).

The cut-off of the RG flow for the effective couplings plays an 
important role in our discussion. It must correspond to a physical 
threshold and thus must be invariant under the various symmetries of the
theory. To clarify this we make a separation in the K\"ahler potential
contributions
\begin{equation}
{\cal K}={\cal K}_S+{\cal K}_T
\end{equation} 
where ${\cal K}_S$ is the $T$ independent  part while  ${\cal K}_T$
changes under the symmetry transformation  (\ref{lab5}).
As an example, in the $Z_3$ heterotic orbifold, ${\cal K}_S$  
stands for the dilaton K\"ahler term $\sim -\ln(S+\overline S)$
and  ${\cal K}_T$  for the K\"ahler term of the  
untwisted moduli  $T,U$ and their higher
order (i.e. one loop and beyond in string coupling)  mixing
with the dilaton and other moduli. A similar 
structure exists for the $Z_3$ orientifold models due to 
an $SL(2,Z)_{T_i}$ symmetry \cite{ibanez1}.
Using this separation we may write eq.(\ref{sugra}) as
\begin{equation}\label{rge}
g_a^{-2}(\mu)= Re f_a+\sigma_a +\frac{-3
T_a(G)}{8\pi^2}\ln\frac{\Lambda e^{{\cal K}_S/2}}{\mu
\left({e^{{\cal K}_S}}/{g_a^2(\mu)}\right)^{1/3}}
+\sum_{r}\frac{T_a^{(r)}}{8\pi^2}\ln 
\frac{\Lambda e^{{\cal K}_S/2}}
{\mu Z_r(\tilde \Lambda,\mu)}
\end{equation}
where we used the notation 
\begin{equation}\label{sigmaa}
\sigma_a= \frac{1}{16 \pi^2} \left[ c_a {\cal K}_T-\sum_{r} 2 T_a^{(r)}
\ln Z_r (\rho, \tilde \Lambda)\right]
\end{equation}
In eq.(\ref{rge}) $\tilde\Lambda$ corresponds to the scale at which
string theory gives the K\"ahler term for the charged matter fields 
in a non-canonically  normalised form  
$K_{matter}=Z_r {\overline Q}_r Q_r \approx
\prod_{i=1}^{3}(T_i+\overline T_i)^{n_r^i} {\overline Q}_r Q_r$
where the approximation ignores higher order string corrections.
We therefore identify the scale $\tilde\Lambda$ with the string 
scale of the theory.

Any variation of $\sigma_a$  induced by changes  
of ${\cal K}_T$ or $Z_r(\rho,\tilde\Lambda)$
under the symmetries of the string theory,
must be cancelled by that of $Re f_a$  (eq.(\ref{lab6})) 
otherwise the predictions for  the low energy physics 
(i.e. $g_a(\mu)$) would not be invariant.
We conclude that the quantity ${\cal F}_a$ defined by
\begin{equation}\label{invariance}
{\cal F}_a \equiv Re f_a +\sigma_a
\end{equation} 
is  invariant under symmetry  
transformations (\ref{lab5}), (\ref{lab4}). Implementing such
symmetries in (\ref{rge}) and (\ref{sigmaa}) to all orders in perturbation 
theory is a very difficult task. This is so because 
${\cal K}_T$ and $Z_r(\rho,\tilde \Lambda)$ (present in $\sigma_a$)
are real functions of the moduli which in general have contributions 
from {\it all  orders} in perturbation theory in the (string) coupling
and we only know their first few perturbative terms. 
Their expressions at the string scale are
\begin{equation}\label{kkk}
{\cal K}_T= {\cal K}^{(0)}_T+{\cal K}^{(1)}_T+\cdots=
-\sum_{i=1}^{3} \ln(T_i+\overline T_i)+{\cal K}^{(1)}_T+\cdots
=-\sum_{i=1}^{3} \ln(T_i+\overline T_i)
\left(1+{\cal O}(g_s^2)+\cdots\right)
\end{equation} 
where only the tree level term is explicitly given.  Higher
order terms  usually mix the dilaton with the rest of moduli fields. 
Here $g_s^2$ is the string coupling. Similarly
\begin{equation}\label{zzz}
Z_{r}=Z^{(0)}_{r}+Z^{(1)}_{r}+\cdots=
\prod_{i=1}^{3}(T_i+\overline T_i)^{n_r^i}+Z^{(1)}_{r}+\cdots
=\prod_{i=1}^{3}(T_i+\overline T_i)^{n_r^i}
\left(1+{\cal O}(g_s^2)+\cdots\right)
\end{equation} 
where $n_r^i$ are the modular weights and again only the tree level 
term is given. Our lack of knowledge of the  higher order
string corrections in the above equations is reflected in the 
accuracy to which $\sigma_a$ and the gauge couplings are 
calculated and this also affects  the unification analysis to follow.
Note that using only string  tree level values 
for $Z$ and ${\cal K}$ only determines the gauge couplings {\it at the
string scale} at one loop level. 

The requirement that ${\cal F}_a$ be invariant under the symmetries
of the theory strongly constrains the form of $f_a$ and $\sigma_a$
in specific string models, with additional implications for the 
unification of the couplings as we now discuss.      
Firstly, this requirement relates the K\"ahler potential for the 
untwisted moduli $T_i$  (present in $\sigma_a$)
to the real part of the holomorphic coupling. Secondly, 
the condition of gauge coupling unification  further relates the latter
to the K\"ahler term for the dilaton as we explicitly show in the next two
sections for specific examples.  This leads to the suggestion 
that there must be a deeper relationship in string theory between
the  K\"ahler term ${\cal K}_T$  for untwisted moduli  and that of 
the dilaton ${\cal K}_S$. Invariance of ${\cal F}_a$   can 
also provide 
a consistency check for string models  with the type of symmetries  
outlined in  the previous section. Other symmetries 
of the K\"ahler potential (for example under $S\rightarrow 1/S$) 
may be implemented in a similar manner.

\vspace{0.5cm}
\subsection{Gauge couplings in $Z_3$ heterotic orbifolds}\label{orbifold}
The heterotic string with gauge group $SO(32)$ compactified on the orbifold 
$T_6/Z_3$, has gauge group $SU(12)\times SO(8)\times U(1)_A$
where $U(1)_A$ is anomalous. This anomaly is universal and is
cancelled by shifts of the dilaton.
A dilaton dependent Fayet Iliopoulos term is then generated and this is
cancelled by v.e.v.'s of the fields $M_{\alpha\beta\gamma}$, the 
blowing-up modes of the orbifold which are present in the 
twisted sector and transform as $(1,1)_{-4}$ under the gauge group.
Additional twisted states $V_{\alpha\beta\gamma}$ 
% (not present on  $Z_3$ orientifold side, see later) 
are present which transform as $V_{\alpha\beta\gamma}=(1,8_s)_{+2}$.
These fields become massive through superpotential couplings
\cite{kakushadze2}. 
% Finally, the closed string spectrum also contains 
% the untwisted moduli $T_i$ and the dilaton. 
For the untwisted string sector there are three families of states 
$Q_a=(12,8_v)_{-1}$ and $\phi_a=(\overline{ 66}, 1)_{+2}$, the
dilaton $S$ and the $T_i$ moduli.  

In heterotic orbifolds  the ``sigma-model''  symmetry\footnote{Only the
discrete subgroup $SL(2,Z)_{T_i}$ survives at the string level.} 
transformation of the $T_i$ moduli is given by
\begin{equation}\label{lab9}
T_i\rightarrow \frac{a_i T_i-i b_i}{i c_i T_i+d_i},\,\,\,\,\,\,
a_i d_i - b_i c_i=1
\end{equation}
The K\"ahler potential of charged matter fields at the string scale 
has the form 
\begin{equation}\label{kmatter}
K_{matter}= Q_{r}^{\dagger}  Z_r  Q_{r}
\end{equation}
Its invariance under $SL(2,R)_{T_i}$ determines $Y(\phi)$ given the
form of $Z_r$, eq.(\ref{zzz}). At the  tree level for $Z$ this implies 
\begin{equation}\label{zzzheterotic}
Z_r= \prod_{i=1}^{3} (T_i+\overline{T_i})^{n_r^i}+\cdots,
\,\,\,\,\,\,\,\,\,\,
Y_r=\prod_{k=1}^{3}(d_i+i c_i T_i)^{n_r^i}
\end{equation}
The string theory  form of the K\"ahler term of the $T$ moduli
is given at the tree level by
\begin{equation}\label{keyt}
{\cal K}_T=-\sum_{i=1}^{3}
\ln(T_i+{\overline T}_i)+\cdots,
\,\,\,\,\,\,\,\,
{\cal J}=\kappa^{-2}\sum_{i=1}^{3} \ln(i c_i T_i +d_i)
\end{equation}
Thus we find 
\begin{equation}
\sigma_a=-\frac{1}{16 \pi^2} \sum_{i=1}^{3} b_a^{' i}\ln
(T_i+{\overline T_i}); \,\,\,\,\,\,\, b_a^{' i}= c_a + \sum_{r}
2 n_r^i T_a^{(r)}
\end{equation}
For the class of $Z_3$ heterotic orbifolds of interest here 
we have \cite{kl2}  $b_a^{' i}\equiv b^{' i}$  i.e. gauge group
independent\footnote{We consider only Kac-Moody level one string
theory.} implying the same  for $\sigma_a$.
The change of $\sigma_a$ under moduli transformation is compensated by 
that of $f_a$ (eq.(\ref{lab6})) which must therefore also contain a gauge
group independent part.  This is indeed the case in
heterotic string theory where $f_a$ is given 
by the universal dilaton  $f_a= S$. Therefore 
\begin{equation}\label{refplussigma}
{\cal F}_a =\frac{1}{2}\left[
S+\overline S -\frac{1}{8 \pi^2} \sum_{i=1}^{3} b^{' i} \ln
(T_i+\overline T_i)\right]\equiv f_o
\end{equation}
and the dilaton is shifted and  ${\cal F}_a$ is $SL(2,Z)_{T_i}$ invariant.
Eqs.(\ref{refplussigma}), (\ref{rge}) together with the requirement  
of the existence of a unified (i.e. gauge group independent)
bare coupling give  $exp(-{\cal K}_S)= {\cal F}_a$
in agreement with the gauge group independence of ${\cal F}_a$. 
This gives 
\begin{equation}\label{fzero}
{\cal K}_S= -\ln \left\{\frac{1}{2} \left[ S+\overline S -\frac{1}{8\pi^2}
\sum_{i=1}^{3} b^{' i} \ln (T_i+ {\overline T}_i)\right]\right\}
\end{equation}
While this result is expected in string theory under the presence of 
$SL(2,R)_{T_i}$ symmetry, we find it interesting that we recovered it
using field theory arguments (anomaly cancellation) and the condition
of unification imposed on the RG flow (\ref{rge}) for the gauge couplings.
The fact that string theory gives a similar 
($SL(2,Z)_{T_i}$ invariant) expression for 
the dilaton potential provides a consistency check of our approach.
Moreover, the definition of ${\cal F}_a$  eq.(\ref{refplussigma})  
includes some dependence of the 
K\"ahler potential for $T$ moduli (\ref{keyt}). From the invariance of
the former under $SL(2,R)_{T_i}$ symmetry together  with the unification
condition relating ${\cal F}_a$ to ${\cal K}_S$   we conclude that the
K\"ahler potentials for   $T$ and $S$ are related at a deeper
level in  string theory where  the unification of the gauge couplings is
respected.

From eq.(\ref{rge}) we find the final form for the RG flow ($\mu\sim M_z$)
\begin{equation}\label{rgfinal}
g_a^{-2}(\mu)= f_0 +\frac{-3 T_a(G)}
{8\pi^2}\ln\frac{\Lambda /\sqrt f_0}{\mu
\left({ f_0^{-1}}/{g_a^2(\mu)}\right)^{1/3}}
+\sum_{r}\frac{T_a^{(r)}}{8\pi^2}\ln \frac{ \Lambda/\sqrt f_0}
{\mu Z_r(\tilde \Lambda,\mu)}
\end{equation} 
We thus identify the unification scale $\Lambda_U$ and the 
RG flow cut-off in (string inspired) supergravity models as 
in eq.(\ref{unifscale}) below. $\Lambda$ is taken equal 
to $M_P$ as the  natural, moduli independent cut-off of RG flow of 
the holomorphic coupling in (\ref{lab2}) \cite{kl2} so
\begin{equation}\label{unifscale}
\Lambda_{U}=\frac{\Lambda}{\sqrt f_o} \approx \frac{M_P}
{\left(S+\overline S - 1/(8\pi^2) \sum_{i=1}^{3}   b^{' i}
\ln (T_i+ \overline T_i)\right)^{1/2}}
\end{equation}
where the approximation sign stands for additional numerical factors
depending on the regularisation scheme \cite{kl4}.
This value of the unification scale, the heterotic
string scale, is slightly
different from that usually quoted 
$M_H= M_P/(S+\overline S)^{1/2}$ \cite{kl2}. 
This difference is due to the mixing at one loop
level between the dilaton and $T$ moduli which we considered, 
so we see eq.(\ref{unifscale}) is the one-loop improved heterotic string
scale.  The numerical effect of the presence of the $T$ dependence in 
this definition is small, since $Re T={\cal} O(1)$
in weakly coupled heterotic strings. 
Our  definition of $\tilde \Lambda$ as the string scale
makes the whole unification  picture at the string scale 
self-consistent since if $\tilde\Lambda=\Lambda_U$, then
$Z_r(\tilde\Lambda,\mu=\Lambda_U)=1$ with the
unified coupling  $g_a^{-2}(\mu=\Lambda_{U})=f_0$ and 
eq.(\ref{rgfinal}) is respected. Note that both the string scale
and the bare coupling are manifestly $SL(2,Z)_{T_i}$ invariant,
as one would expect in a theory with such a symmetry.
Since  $Re T\sim {\cal O}(1)$ we can expand the unified coupling $f_0$  
\begin{equation}\label{f0}
f_o\approx  Re(S+\sum_{i=1}^{3}\epsilon_i T_i)
\end{equation}
which recovers a result of heterotic string 
theory for the structure of the one loop holomorphic coupling. 
If we ignore  the approximations made in 
(\ref{zzzheterotic}) and in (\ref{keyt})
 for ${\cal K}_T$, equations 
(\ref{rgfinal}) and (\ref{unifscale})  are perturbatively exact.
More explicitly, eq.(\ref{rgfinal}) 
uses an input from string theory, the boundary value of $f_0$
for the RG flow below the unification scale, and $f_0$ is known to
one loop order only, however below this scale 
eq.(\ref{rgfinal})  is perturbatively exact.
This concludes our examination of the RG flow 
for the $Z_3$ heterotic orbifold model. The method developed
here applies  generally to the class of 
heterotic string  models without a N=2 sector.

Note that eq.(\ref{rgfinal}) justifies the use of the flat space-time
RG flow, being equivalent to all orders in perturbation theory below
the unification scale to eq.(\ref{mssmlike}).
Moreover it determines the cut-off scale and the bare coupling 
in terms of the fundamental string quantities.

% Finally, there is one interesting aspect already mentioned,
% which we would like to stress.
% Eq.(\ref{rgfinal})  proves that testing unification of the gauge
% couplings in this class of string models is similar to testing this 
% aspect of the RG flow in  MSSM-like models 
% in {\it flat} space-time, eq.(\ref{mssmlike}). This is so because 
% eqs.(\ref{rgfinal}), (\ref{mssmlike}) 
% have a similar form to {\it all orders} in perturbation theory
% {\it below} the unification scale, with the difference that 
% eq.(\ref{rgfinal}) {\it explains} the  origin of the cut-off and of the 
% bare coupling of the running beyond that provided by  
% (\ref{mssmlike}) as simple values
% corresponding to the point where couplings meet.
% For this equivalence to really hold, the more difficult aspect 
% is however how 
% to obtain the appropriate (standard model)
% low energy symmetry group in both cases.

\vspace{0.5cm}
\subsection{Gauge couplings in $Z_3$ orientifolds}\label{orientifold}
The models based on  $Z_3$ orientifolds are similar  to those based on the
heterotic counterpart, the $Z_3$ orbifold. The gauge group is again
$SU(12)\times SO(8)\times U(1)_A$ with the following structure of the
spectrum. The closed string sector contains 
27 twisted moduli $M_{\alpha\beta\gamma}$ corresponding to the
blowing up modes of the associated orbifold and 
with their linear symmetric combination  labelled by $M$.
The closed string spectrum also includes 
the untwisted moduli $T_i$ and the dilaton $S$. The open string sector
has three families of states $Q_a=(12,8_v)_{-1}$, $\phi_a=({\overline
{66}},1)_{+2}$ due to strings stretching between the 9-branes. 
The $U(1)_A$ is again anomalous and the anomaly is this time
non-universal and cancelled by shifting the 
twisted pseudoscalar axions which are in the same chiral multiplets
$M$  as the scalars corresponding to the blowing up modes of the orientifold.
A combination of
the twisted states with  the anomalous vector superfield forms a 
heavy vector multiplet, which after decoupling at a high
scale\footnote{This is of the order of the string scale, 
just as in the case of the heterotic orbifold \cite{klein}.}
 ``leaves'' a  $SU(12)\times SO(8)$ gauge group, just  as in 
the  $Z_3$ orbifold model.
We see that  below this scale there is a match of the spectrum and of the 
gauge groups of $Z_3$ orbifold and $Z_3$ orientifold models 
\cite{nilles0,lavignac}. 
This suggests the  possibility that $Z_3$ orientifold is a good
candidate  to be the dual model of the $Z_3$ orbifold
\cite{angelantonj1,angelantonj}, motivating the suggestion that the $Z_3$ 
orientifold model has a {\it non-anomalous} 
$SL(2,Z)_{T_i}$ symmetry, just like its
heterotic dual. As we have seen the $SL(2,Z)_{T_i}$ anomalies are
important in determining the structure of the gauge couplings running.
Here we determine the anomaly structure in the $Z_3$ orientifolds, 
although some of our results could apply to more general cases.
Since in the heterotic case such anomalies are cancelled,
the proposed orientifold-orbifold duality implies that these 
anomalies are also cancelled \cite{ibanez1} 
for $Z_3$ and $Z_7$ orientifolds. The duality 
symmetry which prompted the study of the cancellation of these
anomalies in type IIB orientifolds was investigated beyond tree
level  in  \cite{nilles}.  However, this analysis
was based on the linear - chiral multiplet transformation \cite{antoniadis}
which was assumed to hold at one loop level and this was proved using only
the tree-level string scale in the linear basis for the gauge couplings 
eq.(\ref{eq1}),  rather than its one-loop improved value.
This tree level definition of the string scale  is not invariant 
under the symmetry transformation of $T_i$'s and from  eq.(\ref{eq1})
this  means that low energy physics  is not invariant either.
In  the heterotic case the string scale 
changes at one loop\footnote{For this see previous section.}, giving  an
invariant form (\ref{unifscale}). We expect something similar 
should apply to the orientifold case as well.
 We therefore 
conclude that the  linear-chiral multiplet duality relation  may prove to be 
more complicated than assumed and that
 the two models $Z_3$ orbifold/orientifold
could still be dual to each other. We do not make explicit use of this
 duality, but we will  later 
discuss its compatibility with our results
in the effective field theory approach where all states considered 
are in the chiral basis. From this (``string based'') effective 
theory point of view  it is useful to investigate
the phenomenological consequences for the running couplings
and 
% clarify the scenario of ``mirage unification'' \cite{ibanez2,ibanez3,halyo} 
discuss the unification of the gauge couplings
in the presence of  the $SL(2,Z)_{T_i}$ symmetry transformation
\footnote{Note that  transformation  (\ref{lab9})
is not a $T$  duality transformation in type I vacua. The latter
exchanges different type of D branes, the three $T_i$'s
and the dilaton \cite{ibanez5}. Clearly, transformation (\ref{lab9})
is not of this type. Further, in the orientifold model we examine with
the proposed symmetry only D9 branes are  present.} of the $T_i$ 
moduli, eq.(\ref{lab9}). 

As in the heterotic case we take as input from the $Z_3$ orientifold
string theory the K\"ahler terms for the untwisted moduli $T$
\begin{equation}\label{lab8}
{\cal K}_{T}=-\kappa^2{\tilde\kappa}^{-2}
\sum_{i=1}^{3}\ln(T_i+{\overline T_i})+\cdots
\end{equation}
where ${\tilde\kappa}^{-2}$ is  the coefficient in front of the 
K\"ahler potential as given after compactification
in string theory for $Z_3$ orientifold
while $\kappa^2$ is due to  definition (\ref{lab2}).
In this class of orientifolds we also have 
a K\"ahler term for the charged matter fields of the form 
(\ref{kmatter}) with (\ref{lab5}), (\ref{lab4})   respected.
Under the combined effect of eqs.(\ref{lab5}), (\ref{lab4})
we find from (\ref{sigmaa})
\begin{equation}\label{lab11}
\sigma_a=-\frac{1}{16 \pi^2} \sum_{i=1}^{3} b_a^{' i}
\ln(T_i+{\overline T}_i)
\;\;\;\;\;\;\;
b_a^{' i}=\kappa^2 {\tilde \kappa}^{-2} 
c_a+\sum_{r} 2 T_a^{(r)} n_r^i
\end{equation}
Compatibility with the  string calculation for $b_a^{' i}$ for the $Z_3$
orientifold \cite{ibanez1} requires that 
$\epsilon \equiv \kappa^2 {\tilde \kappa}^{-2}$
be equal to 1. 
Unlike the case of the $Z_3$ heterotic orbifold, $b_a^{\prime i}$ 
in the $Z_3$ orientifold is gauge group dependent due to the different 
spectrum content. We have $b_a^{' i}=b_a/3$ where $b_a$ is 
the one loop beta function of the associated gauge group 
$SU(12)\times SO(8)$.
This is because in the $Z_3$ orientifold  there is no  counterpart 
of the states $V_{\alpha\beta\gamma}$ of the $Z_3$ heterotic orbifold.

The requirement that ${\cal F}_a$, eq.(\ref{invariance}) 
be invariant requires that $f_a$ have indeed a non-zero
gauge group dependent part and, if unification exists,
a gauge group independent part as well. Therefore 
$f_a=S +\sum_{k} s_{ak} M_k$ which establishes the 
structure of the holomorphic coupling for this model.
For the gauge group dependent  part
the presence in $f_a$ of the fields $M_k$ is required  
on pure field theory grounds to cancel the gauge group dependent 
variation of $\sigma_a$ 
under $SL(2,Z)_{T_i}$ (anomaly cancellation). 
Further, we know from the string theory that the fields $M_k$
are indeed singlets under the gauge  group of the orientifold.
The gauge independent part of $f_a$ cannot be proportional
to  moduli other than $S$ like for example $T_i$'s 
because the latter would transform non-linearly 
under (\ref{lab9}) and invariance of ${\cal F}_a$ 
(\ref{invariance}) would not be respected. 
 Further, we may consider that
$S$ (identified as the dilaton $S$)
is not shifted under  transformation\footnote{This is 
an input from string theory \cite{ibanez1}. See later for discussion 
on this point.}  (\ref{lab9}).
 These results  are indeed in agreement with type II B
$Z_N$  models, $N=odd$  where the holomorphic coupling $f_a$
in the presence of 9-branes only is given by 
\begin{equation}\label{lab11prime}
f_a=S+\sum_{k=1}^{(N-1)/2}s_{a k}M_k
\end{equation}
with $M_k$ twisted moduli (chiral basis). 
For $Z_3$ orientifold there is only one field $M_k\equiv M$. 
From now on, whenever possible we  will consider the more 
general  case of $Z_N$ 
orientifold and in the results  for  the $Z_3$ case we drop the index
$k$. We then find from (\ref{invariance}), (\ref{lab6}) that 
the twisted moduli transform according to
\begin{equation}\label{lab12}
M_k\rightarrow M_k^\prime = M_k-\frac{1}{8\pi^2}
\sum_{i=1}^{3}\delta_{i}^{k} \ln (d_i+i c_i T_i )
\end{equation}
where coefficients $s_{ak}$ satisfy  the relationship 
\begin{equation}\label{lab13}
\sum_{k} s_{ak}\delta_i^k=b_a^{\prime i}
\end{equation}
The result  (\ref{lab12}) recovers 
the initial string-based suggestion of
\cite{ibanez1} that anomalies can be cancelled by the transformation
of the twisted moduli (\ref{lab12}) for $Z_3$ and $Z_7$ 
orientifolds. However, this is only true  if $\epsilon=1$ 
when  condition (\ref{lab12})
is identical to that proposed in string theory \cite{ibanez1}.
This therefore fixes the coefficient in front of the  
K\"ahler term for the moduli, eq.(\ref{lab8}) for the agreement 
of field-string theory  to hold, and the situation is then very 
similar to that of the heterotic string.
However eq.(\ref{lab12})   assumes that the
dilaton is inert under $SL(2,Z)_{T_i}$ symmetry 
and plays no role in anomaly
cancellation. In principle the dilaton could
be shifted by a universal (gauge group independent) term as well
compensated by an opposite shifting of the twisted moduli part of
$f_a$, but this is not allowed in string theory \cite{nilles,ibanez1}.
Finally, we note that
$Re M_k$ cannot represent  a vacuum state invariant 
under $SL(2,Z)_{T_i}$ and cannot be set to zero
% \footnote{This seems to  contradict  \cite{ibanez3}}, 
as in such a case anomaly cancellation and invariance of ${\cal F}_a$
eq.(\ref{invariance}) would not be respected, and low energy
physics would not be invariant under the transformation 
of $T_i$ eq.(\ref{lab9}).

For the  $Z_3$ orientifold  with a single field $M$ present,
eq.(\ref{lab13}) shows that  $s_{ak}\equiv s_a \propto b_a^{' i}=b_a/3$
on pure anomaly cancellation grounds (also $\delta_i=6$). 
The proportionality of $s_{a}$
coefficient to the one loop beta function is again in agreement with
explicit string calculations of these coefficients 
\cite{antoniadis,ibanez1}. We would like to stress that unlike the
string calculation \cite{antoniadis}, this proportionality 
is here  a consequence of imposing the anomaly
cancellation under the conjectured $SL(2,Z)_{T_i}$ symmetry. 
This provides a check for our approach,  circumstantial 
evidence  for the presence of this symmetry {\it at string level} 
and further motivation for studying its implications.

We can now proceed to investigate the RG flow for the $Z_3$
orientifold and the unification of the gauge couplings.
Their values at the string scale $\tilde \Lambda$  are given by
(using (\ref{rge}),  (\ref{kkk}),  (\ref{zzz}),
(\ref{lab11prime}), (\ref{lab13}))
\begin{eqnarray}\label{rg2p}
g_a^{-2}(\tilde \Lambda)&=& Re f_a +\frac{-3 T_a(G)}{8\pi^2}
\ln\frac{\Lambda  e^{{\cal K}/2}} {\tilde\Lambda 
\left({e^{{\cal K}}}/{g_a^2(\tilde \Lambda)}\right)^{1/3}}
+\sum_{r}\frac{T_a^{(r)}}{8\pi^2}\ln 
\frac{\Lambda e^{{\cal K}/2} }{\tilde\Lambda Z_r(\rho,\tilde\Lambda)}\\
 &=&    Re S +\frac{-3 T_a(G)}{8\pi^2}
\ln\frac{\Lambda  e^{{{\cal K}_S}/2} e^{\cal G} } {\tilde\Lambda 
\left({e^{{\cal K}_S}}/
{g_a^2(\tilde \Lambda )}\right)^{1/3}}
+\sum_{r}\frac{T_a^{(r)}}{8\pi^2}\ln 
\frac{ \Lambda e^{{{\cal K}_S}/2} e^{\cal G}} 
{\tilde\Lambda {\cal Z}_{string} % \{1+{\cal O}(g_s^2)\}
}
+\frac{c_a}{16 \pi^2} 
\left[{\cal K}_T^{(1)}+\cdots \right]\nonumber\\
\label{secondpart}
\end{eqnarray}
where we used the notation
\begin{equation}
{\cal Z}_{string} =\{1+{\cal O}(g_s^2)\},\,\,\,\,\,\,\,\,
{\cal G}= \frac{2\pi^2}{9} \left\{M+{\overline M} -
\frac{1}{8\pi^2}\sum_{i=1}^{3} \delta_i \ln (T_i+{\overline {T_i}})\right\}
\end{equation}
while the effective couplings at low energy scales ($\mu\sim M_z$) 
are given by
\begin{equation}\label{rg1}
g_a^{-2}(\mu)= g_a^{-2}(\tilde \Lambda) 
+\frac{-3 T_a(G)}{8\pi^2}\ln\frac{\tilde \Lambda}
{\mu \left({g_a^2(\tilde \Lambda)}/
{g_a^2(\mu)}\right)^{1/3}}+\sum_{r}\frac{T_a^{(r)}}{8\pi^2}
\ln \frac{\tilde \Lambda}{\mu Z_r(\tilde \Lambda,\mu)}
\end{equation}
The low energy physics represented by $g_a^2(\mu\sim M_z)$ 
is indeed invariant under 
$SL(2,Z)_{T_i}$ transformation since ${\cal G}$ itself 
is  invariant. This fact can be seen explicitly 
by adding 
eqs.(\ref{secondpart}) and (\ref{rg1}) and ignoring the extra terms
due to higher order (one-loop) string corrections to $Z$ and ${\cal K}_T$
at/above the string scale\footnote{One-loop  order (and beyond)
string-induced corrections to $Z$ and ${\cal K}_T$,  
normalised to their tree level values, induce two-loop-like (and beyond)
terms in the RG flow for  the effective gauge couplings,
due to physics at/above the string scale. These corrections
have the structure similar to but distinct from
that of field theory two loop terms.}
 (denoted in (\ref{secondpart}) 
by ${\cal Z}_{string}$ 
and ${\cal K}_T^{(1)}\sim {\cal O}(g_s^2)$ 
respectively). Such extra terms  would affect some of the two loop 
corrections (of string origin!) for the gauge couplings. 
Even though the string scale $\tilde\Lambda$ and consequently 
$g_a(\tilde \Lambda)$ are not  invariant under an $SL(2,Z)_{T_i}$
symmetry, the low energy physics $(g_a(\mu), \,\, \mu\sim M_z)$
is still invariant (in this approximation), as should be the case. 
To see explicitly if this true beyond this approximation we would
need an explicit string calculation of ${\cal Z}_{string}$ 
and ${\cal K}_T^{(1)}$.

If the couplings unify beyond one-loop level one sees from 
eq.(\ref{secondpart}) that ${\cal K}_S=-\ln(S+\overline S)$.
This identification recovers the 
structure for the K\"ahler potential  given by a string calculation.
In this case the couplings unify at the scale 
$\Lambda'$ where
\begin{equation}\label{lambdaprime}
\Lambda'=\Lambda e^{ {{\cal K}_S}/2} e^{\cal G}
\end{equation}
Up to the terms ${\cal O}(g_s^2)$ the unified coupling
is ``fixed'' by the dilaton alone, $g^{-2}(\Lambda')= Re S$
rather than in combination with the twisted moduli. 
This is supported by the fact that (unlike $M_k$)  $S$ 
is not involved in anomaly cancellation and is invariant under the
$SL(2,Z)_{T_i}$.

\subsubsection{\it The  unification scale
and the linear-chiral multiplet relation}

The result (\ref{lambdaprime}) for the unification scale 
certainly requires  some discussion. 
From the supergravity point of view it has been
argued that the scale $\Lambda$ entering the definition of $\Lambda'$ 
eq.(\ref{lambdaprime}), should be identified with the Planck mass, $M_P$. 
In the heterotic string this leads to the conclusion 
that the couplings unify at the (one-loop improved) heterotic string scale 
given by eq.(\ref{unifscale}).
For the effective supergravity RG flow eqs.(\ref{rg2p}),(\ref{secondpart})
applied to the $Z_3$ orientifold model the important point is that
$\Lambda$ does not bring any
moduli dependence in definition (\ref{lambdaprime}).

% It is not clear to us that $\Lambda=M_P$
% for the effective supergravity RG flow
% eqs.(\ref{rg2p}),(\ref{secondpart})
% applied to the $Z_3$ orientifold model. 
% However, for the following discussion we consider
% that this is indeed the case and thus $\Lambda$ does not bring any
% moduli dependence in (\ref{lambdaprime}).

If the unification scale is larger than the string scale, 
$\Lambda'> \tilde\Lambda$, the  structure of eq.(\ref{secondpart}) 
mimics the field theory running of the gauge 
couplings above  $\tilde\Lambda$.
This is not RG flow in the field theory
sense because the radiative corrections are of string origin (moduli 
contributions) 
which do not have a field theory correspondence.  
The effect of these moduli contributions 
is to renormalise the wavefunctions of the 
gauge sector (through the presence of $exp({\cal K}_S/2)$ on the
r.h.s. of (\ref{secondpart})) and those of the matter sector 
(through the presence of $Z_{string}\equiv \{1+{\cal O}(g_s^2)\}$ 
on\footnote{The origin of these corrections is in eq.(\ref{zzz}).}
the r.h.s. of (\ref{secondpart})).
This means that there may be a stage of ``mirage unification'' 
\cite{ibanez4prime} at $\Lambda'$  induced by string effects only.
It is possible however that  $\Lambda'=\tilde\Lambda$ and 
unification actually takes place at the string scale
$\tilde\Lambda$. To distinguish between these two cases of
unification one needs to know the exact value of ${\cal G}$. 

% The result (\ref{lambdaprime}) for the unification scale 
% certainly requires  some discussion. 

The value of ${\cal G}$ can be fixed by the Fayet Iliopoulos
mechanism\footnote{For
more on Fayet Iliopoulos mechanism in type IIB orientifolds see 
\cite{cvetic,nilles0}.} in the following way. In general the D term in
the Lagrangian contains in addition to the Fayet 
Iliopoulos term, proportional to ${\cal G}$ \cite{klein}, a contribution 
from  fields charged under the anomalous
$U(1)_A$. However at least for $Z_3$ orientifold model we 
considered, fields charged
under $U(1)_A$ are also charged under the non-Abelian group. A
non-vanishing v.e.v. of these fields would therefore trigger a
breaking of the non-Abelian symmetry group factors.
In the following we will discuss two possibilities
corresponding to whether we insist on the presence or absence of
the full non-Abelian gauge symmetry of the model. 

If we  insist that the full non-Abelian symmetry be unbroken,
then the D term in the Lagrangian depends only on the 
contribution given by ${\cal G}$.
This leads to the conclusion that the
vanishing  of the $D$ term contribution in the Lagrangian (to preserve
supersymmetry) implies the 
vanishing of  ${\cal G}$ as well \cite{klein} in the orientifold 
limit \cite{klein}.
To illustrate this  case\footnote{In the following we will consider
the more general case of $Z_N$ orientifolds, N odd. For $Z_3$
orientifold one must suppress the indices $k$.},  consider the 
twisted moduli   K\"ahler potential $K(M_k, \overline M_k)$
which  must be invariant  under the $SL(2,Z)_{T_i}$ symmetry 
and the anomalous $U(1)_A$, conditions  which  lead  to  the following 
change of its original form
%\footnote{
%However it is possible to have additional, $SL(2,Z)_{T_i}$ 
%invariant terms as arguments of $K_M$, leading  to a 
%different, non-zero value for ${\cal G}$, see later.} 
\begin{equation}\label{lab14}
K_M(M_k,{\overline M_k})\equiv K_M(M_k+{\overline M_k})\rightarrow 
K_M(M_k+{\overline M_k}-\delta_k V_A-\frac{1}{8\pi^2}
\sum_{i=1}^{3}\delta_i^k \ln(T_i+{\overline T_i}))
\end{equation}
This contribution is not present in the running couplings
(eq.(\ref{rg2p})) because it is considered a higher order contribution
in the moduli. However it does control the value of ${\cal G}_k$ (${\cal G}$
for $Z_3$) through the
Fayet Iliopoulos mechanism \cite{klein}. 
This requires, in the case of unbroken non-Abelian symmetry of $Z_N$ 
orientifold
\begin{equation}\label{lab15}
\sum_{k=1}^{(N-1)/2}\delta_k 
\frac{\partial{K_M}}{\partial{M_k}}
\bigg\vert_{V=0;\\ \theta=\overline\theta=0}=0
\end{equation}\label{lab16}
and therefore a sufficient (and  necessary for $Z_3$ orientifold) 
condition is that 
\begin{equation}\label{lab17}
\frac{\partial{K_M}}{\partial{M_k}}
\bigg\vert_{V=0; \theta={\overline\theta}=0}=0
\end{equation}
As was shown in \cite{klein} this corresponds (for nonsingular
potential) to the condition
\begin{equation}\label{GK}
{\cal G}_k\equiv 
\frac{2\pi^2}{9} \left\{M_k+{\overline M_k} -
\frac{1}{8\pi^2}\sum_{i=1}^{3} \delta_i^k 
\ln (T_i+{\overline {T_i}})\right\}=0
\end{equation}
For the case of the $Z_3$ orientifold (${\cal G}_k\equiv {\cal G})$)
we thus find that the unification scale $\Lambda'=\Lambda e^{{\cal
K}_S/2}$. 
However, it is possible that $K_M$ has additional $SL(2,Z)_{T_i}$
invariant contributions to its argument, eq.(\ref{lab14}), leading to
non-zero value for ${\cal G}_k$ (${\cal G}$ for $Z_3$) (and which  
may give a large value for $\Lambda^\prime$).
Moreover, the result of eq.(\ref{GK}) for ${\cal G}_k$ (${\cal G}$) 
is questionable for two additional reasons.

Apparently, an unification scale 
of value $\Lambda e^{{\cal K}_S/2}=M_P/(S+{\overline S})^{1/2}$ 
would  be very encouraging from a
phenomenological point of view and would preserve some similarities 
with the heterotic case. 
In fact an explicit string calculation \cite{antoniadis}
has shown that in orientifold models the contribution of the N=2
sector indeed extends beyond the string scale up to the winding mode
scale, which could be close to the Planck scale. However in the
present case there does not appear to be a physical threshold
associated with this,
so this possibility appears  implausible.

Secondly, the  vanishing of ${\cal G}_k$ is not compatible with
the linear-chiral multiplet duality relation \cite{antoniadis}
which relates eqs.(\ref{eq1}), (\ref{rg2p})
\begin{equation}\label{duality}
2 m_k= \underbrace{M_k+\overline M_k - \frac{1}{8\pi^2}\sum_{i=1}^{3} 
\delta_i^k\ln(T_i+{\overline T}_i)}_{\sim {\cal G}_k}
+ \frac{b_k}{8\pi^2}\ln\left[
\frac{\prod_{i=1}^{3} Re T_i}{Re S}\right]^{1/2}
\end{equation}
with $b_a=\sum_{k} b_k s_{ak}$, ($b_k\equiv b=18$ for $Z_3$).
The origin of this disagreement was highlighted in the Introduction,
and is due to the $T_i$ dependence of the tree level string scale in $Z_N$ 
orientifold models,  
$M_I^2=2 M_P^2 (S+\overline S)^{-1/2}
\prod_{i=1}^{3}(T_i+\overline T_i)^{-1/2}$.
This extra $T_i$ dependence brought in solely
by the string scale definition   is manifest 
in the linear-chiral multiplet relation, the last term on the r.h.s.
of (\ref{duality}). The $T_i$ dependence of this term  is not present
in the linear basis of the Lagrangian   leading to 
${\cal G}_k=0$ in ref.\cite{klein} in the orientifold 
limit, $m_k\rightarrow 0$.  
One could attempt to preserve the multiplet duality 
eq.(\ref{duality}), and for this the Lagrangian of linear multiplets 
of  \cite{klein} should contain an additional term \cite{nilles}, 
given by
\begin{equation}\label{extraterm}
\Delta {\cal L}_2=\frac{1}{16 \pi^2}\sum_{k=1}^{(N-1)/2} 
b_k {\hat L_k}\ln \left[
{\hat L_k} \prod_{i=1}^{3} \left(T_i+{\overline T}_i\right)\right]
\end{equation} 
in the notation of \cite{nilles} with $\hat L_k$ standing for the linear
multiplet basis.
This brings in the third term in the rhs of (\ref{duality})
and leads to ${\cal G}_k\not = 0$ in the orientifold limit,
$m_k\rightarrow 0$.
However, the value of ${\cal G}_k$ in this limit
 as given by (\ref{duality}) is not 
$SL(2,Z)_{T_I}$ invariant, contrary to what we already established,
eq.(\ref{lab12}).
This is because the last term in (\ref{duality}) is not invariant and
ruins the   anomaly  cancellation  of the model as observed in \cite{nilles}.
All these difficulties are caused by the fact that the
tree level definition of the string scale, $M_I$
responsible for the last term on the rhs of
(\ref{duality}) is not $SL(2,Z)_{T_i}$ invariant. As we observed after
eq.(\ref{eq1}), for the same reason $g_a(\mu\sim M_z)$ in
(\ref{eq1})  is not invariant under $SL(2,Z)_{T_i}$. 

It is possible that the tree level string scale $M_I$
is made $SL(2,Z)_{T_i}$  invariant by
one loop corrections due to the moduli, similarly to the heterotic case,
eq.(\ref{unifscale}); this should actually be the case
since this scale is a physical one and must be invariant if the 
symmetry $SL(2,Z)_{T_i}$  is present. In this case the 
invariance of $g_a(\mu\sim M_z)$ in (\ref{eq1}) is assured, if we
replace $M_I$ by $\tilde\Lambda$, where the latter is assumed invariant.
As $S$ does not play any role
in anomaly cancellation in $Z_N$ orientifold, and is therefore fixed
(to all orders) this means that the $T_i$ dependence of the one loop improved 
value of the string scale must be $SL(2,Z)_{T_i}$ invariant. One possibility 
is that this value is given by 
\begin{equation}
\tilde\Lambda=M_I \left[\gamma(T_1,T_2,T_3)\right]^{-1/2}
\end{equation}
where  
$\gamma(T_1,T_2,T_3)$ is such as to keep $\tilde\Lambda$ invariant under the
transformation of $T_i$. An example is  
$\gamma(T_1,T_2,T_3)=\prod_{i=1}^{3}|\eta(iT_i)|^2$. 
However, one needs a confirmation of this,
based on a string calculation of the radiative corrections to 
eq.(\ref{eq1})\footnote{This may be difficult to check. 
Unlike the heterotic case where the sigma model symmetry
has an underlying string equivalent (T duality), it is not clear 
if this symmetry holds exactly in  type I string \cite{ibanez1}. 
The heterotic-type I
duality in 10D would suggest this symmetry does not survive in 
type I perturbation theory and therefore the
appearance of the term $\ln\eta(iT)$  
in (\ref{eq1}) would be of  non-perturbative origin.}. 
The new linear-chiral multiplet
duality relating  (\ref{rg2p}) to (\ref{eq1}) (with $M_I$ replaced by
$\tilde\Lambda$ to keep $g_a(\mu\sim M_z)$ $SL(2,Z)_{T_i}$ invariant)
is in this case similar to (\ref{duality}) but with an additional factor 
$\gamma(T_1,T_2,T_3)$ under the last log in (\ref{duality}).
As the linear-chiral relation in the orientifold limit (eq.(\ref{eq1})
with $M_I\rightarrow \tilde\Lambda$, $m_k\rightarrow 0$)
essentially imposes unification at the
``new'' string scale $\tilde\Lambda$,  ${\cal G}$ will be 
such that $\Lambda'=\tilde\Lambda$, and there is no ``mirage''
unification.
We conclude with the remark that in the absence of 
additional string corrections to the tree level string scale $M_I$
in eq.(\ref{eq1}), of the nature
discussed above, the $SL(2,Z)_{T_i}$ symmetry seems implausible, casting
doubts on the $Z_3$ orbifold/orientifold duality.

So far we have considered that the full non-Abelian gauge symmetry of the 
$Z_3$ orientifold model was preserved. However it is possible that this
symmetry  will be broken  if matter fields charged under the $U(1)_A$
symmetry, also charged under  the non-Abelian gauge group  are
present and develop a v.e.v. This is indeed possible for the $Z_3$
orientifold model considered.
In this situation the D term in the
Lagrangian  contains
additional contributions from the charged matter fields.
The condition of preserving supersymmetry
(the vanishing of the D term) will then lead to a non-vanishing value
for ${\cal G}$ different from the case discussed above, due to the
additional v.e.v. contributions.
These  contributions lead to the  ``mirage unification''
scenario in the sense that $\Lambda'$ may be situated above the string
scale $\tilde\Lambda$. The ``mirage'' unification
would not mean an effective ``running'' of the couplings above the 
string scale, but simply a change from $\tilde\Lambda$ to a possibly
higher scale due to the presence of the v.e.v. of fields charged 
under $U(1)_A$ and the initial non-Abelian group.

% These observations prevent us from giving a definite answer 
% about the exact value of ${\cal G}$  and therefore
% of the unification  scale, $\Lambda'$
% consistent with the linear-chiral multiplet duality relation.
% A further investigation  of this relation, in the presence of the
% proposed symmetry  is necessary to  fix the value of ${\cal G}$
% in the orientifold limit.

To conclude, the most likely possibility we 
envisage is that, if $SL(2,Z)_{T_i}$ 
symmetry is indeed present, the couplings unify at the string scale
($\tilde\Lambda$),
if the initial non-Abelian group is unbroken, but  the linear-chiral 
multiplet relation must be changed.
It is also possible that the gauge group is broken to a non-Abelian subgroup
so ${\cal G}$ receives additional corrections due to the v.e.v. 
of the fields involved in the gauge symmetry breaking. In this case it
is possible that $\Lambda'$  be situated above the string scale due
to the v.e.v. which modify the value of ${\cal G}$ from the previous case.
(Of course, an alternative possibility is that the symmetry $SL(2,Z)_{T_i}$ 
is not present at all  for the $Z_3$ orientifold.)

\section{Conclusions}
In this paper we investigated the RG flow in two string inspired models 
based on the $Z_3$ orbifold and orientifold respectively. The RG flow 
in Supergravity proves to be a strong tool for exploring various
aspects of these models. In $Z_3$ orbifold based models we have shown that
RG flow together with the requirement of gauge coupling
unification provides information on the 
structure of the K\"ahler potential for the dilaton and the values of
the unified coupling/unification scale. In the case of $Z_3$
orientifold model we have
shown that anomaly cancellation requires the presence of additional
fields $M$ to cancel the  anomalies induced by the $SL(2, Z)_{T_i}$
symmetry. Moreover, their contribution to the gauge couplings comes 
with a coefficient proportional to the one loop beta function, just 
as is found in string calculations. Finally, the invariance of the 
dilaton under the proposed symmetry and the presence of
unification  suggests that the dilaton can act as the bare coupling 
on its own (rather than in linear combination with the twisted moduli),
at a scale  of order 
 $\Lambda/(S+\overline S)^{1/2} e^{\cal G}$. 
The  exact value of this scale depends on the value of ${\cal G}$
leading to two possibilities. The first of these preserves unification
at the string scale, but the linear-chiral multiplet duality must be
changed. The second possibility may realise a ``mirage'' 
unification due to
v.e.v. of charged matter fields breaking the initial group to a
non-Abelian subgroup. These v.e.v.'s  bring additional corrections 
to ${\cal G}$ from the previous case,  and thus  $\Lambda'$ may be
larger  than $\tilde\Lambda$.

% The precise value of the latter strongly depends on the Fayet Iliopoulos
% mechanism. This mechanism must be implemented 
% such as the linear - chiral duality 
% condition is itself  invariant under $SL(2,Z)_{T_i}$. 
% This should be so because the  low energy physics ($g_a(\mu\sim M_z)$) 
% must be invariant under such transformation 
% in both linear  and chiral basis, and therefore the same
% should apply to the  (linear-chiral) relation which relates them.

\section{Acknowledgements}
The authors would like to thank E. Dudas,  L.E. Ib\'a\~nez,
M. Klein, Z. Lalak, S. Lavignac and H.P.~Nilles for 
very useful discussions.
D.G. acknowledges the financial support from the part of 
the University of  Oxford, Leverhulme Trust research grant. 
This research is supported in part by the 
EEC under TMR contract ERBFMRX-CT96-0090.

\end{document}